\documentstyle[epsfig]{aipproc}
\newcommand{\STRUT}{\rule{0in}{2.5ex}}
\font\vgr=msam10
\newfam\symfam \def\bmit{\fam\symfam\vgr}
\textfont\symfam=\vgr
\mathchardef\fsquare="7104
\mathchardef\fstar="7146
\mathchardef\fdiamond="7107
\def\bdia{$\bmit\fdiamond$}
\def\bstar{$\bmit\fstar$}
\def\bsq{$\bmit\fsquare$}
\def\qq{\hbox{~{\lower -3pt\hbox{$<$}}\hskip -6pt\raise
  -3pt\hbox{${}_\sim$}}}
\begin {document}

\title{Twenty-Five Years of Progress in the\\Three-Nucleon Problem}

\author{J.\ L.\ Friar}
\address{Theoretical Division\\ Los Alamos National Laboratory\\
Los Alamos, New Mexico, 87545}

\maketitle

\begin{abstract}
Twenty-five years ago the International Few-Body Conference was held in Quebec
City.  It became very clear at that meeting that the theoretical situation
concerning the $^3$He and $^3$H ground states was confused. A lack of
computational power prevented converged brute-force solutions of the Faddeev or
Schr\"{o}dinger equations, both for bound and continuum states of the
three-nucleon systems.  Pushed by experimental programs at Bates and elsewhere
and facilitated by the rapid growth of computational power, converged solutions
were finally achieved about a decade later.  Twenty-five years ago the first
three-nucleon force based on chiral-symmetry considerations was produced. Since
then this symmetry has been our guiding principle in constructing three-nucleon
forces and, more recently, nucleon-nucleon forces.  We are finally nearing an
understanding of the common ingredients used in constructing both types of
forces.  I will discuss these and other issues involving the few-nucleon systems
and attempt to define the current state-of-the-art.
\end{abstract}

\section*{Introduction}

The purview of my talk is progress that has been made in our understanding of
the three-nucleon systems and of the dynamics that underlies that understanding.
My emphasis will be on the theoretical side. My reference point in time is 1974,
the date when Bates first delivered beam for an experiment. I will survey that
progress by referring to two other significant events that occurred in 1974. 
One of these is personal:  I attended the International Few-Body Conference held
that year in Quebec City, Canada\cite{Quebec}. The second event is the genesis
in that year of three-nucleon forces (3NFs) based on chiral-symmetry
considerations\cite{Yang}.

On a personal note it is always a pleasure to return to MIT, where I was a
postdoc.  Looking back at my work during that period, I find that almost
everything dealt with electron scattering, a result of the influence of Bates on
the young theorists in the Center for Theoretical Physics.  Part of that work
involved relativistic corrections to the charge densities of few-nucleon
systems, and that motivated my attendance at the Quebec meeting.

There are basically three reasons why three-nucleon physics has become a
subfield in its own right.  The first is that the trinucleons are rich,
nontrivial, and ``simple'' nuclear systems, and understanding their properties
is a minimal criterion for success in this area. The word ``simple'' in this
context means that we are capable of performing the very difficult calculations
of three-nucleon properties. Indeed, in recent years we have not only succeeded
in performing these calculations, but have achieved an understanding of most of
the basic trinucleon properties\cite{joe,walter}.

The second reason is the classic and original goal of the field:  using these
systems to sort out and refine our understanding of the nuclear force.  This is
the most important remaining aspect of the problem, which has been greatly aided
in recent years by chiral perturbation theory ($\chi$PT).  Much of our
theoretical and experimental attention has been directed at 3NFs, because
trinucleon properties show relatively little sensitivity to the details of
modern $N$-$N$ forces. Our remaining problems (though few) are likely due to our
lack of understanding of 3NFs \cite{3NF}.

Finally, the lovely techniques used in this field are fun to work with, and this
attraction has seduced two generations of theorists.  Our efforts have led to
the very successful application of few-body methods to heavier systems, which
goes far beyond even the dreams of 1974, as shown at this symposium by Vijay
Pandharipande.

My strongest impressions of the Quebec meeting are that the field was in a state
of confusion.  Many calculational techniques were in use, each giving a
different answer to the same problem, the $^3$H bound-state energy. Faddeev
methods, hyperspherical expansions, variational bounds, and separable
approximations all had their practitioners\cite{Quebec}. There was a 10-20\%
uncertainty ($\sim$ 1-2 MeV) in the $^3$H binding energy, implying that most (in
retrospect, all) of the calculations were not converged. The situation was
similar with respect to scattering calculations. In order to achieve convergence
one requires brute-force computational resources on a scale that would not be
available for another decade.

\section*{Nuclear Forces}

The genesis of the computational problem is the spin of the nucleon. Contrary to
much folklore, nuclear physics is difficult not because the force is complicated
(in shape), but because it is complex (i.e., it has many components). The origin
of the problem is the spin and parity of the pion:  $J^\pi = 0^-$. The
$\pi$-nucleon vertex must have a complementary pseudoscalar structure in order
to conserve angular momentum and parity, and the dominant form $(\sim 1^+ \cdot
1^-)$ is $\vec{\sigma}_N \cdot \vec{q}$, where $\vec{\sigma}_N$ is the nucleon
(Pauli) spin and $\vec{q}$ is the pion momentum.  This leads immediately to a
tensor component of the force (part of the one-pion-exchange potential (OPEP)),
which dominates interactions in few-nucleon systems.  Indeed, $\langle V_{\rm
OPEP}\rangle$ is roughly 75\% of the total potential energy.  This spin
dependence, together with isospin dependence, accounts for the complexity.  Each
nucleon has $2\cdot2=4$ spin-isospin components, implying that there are roughly
$(4)^2 = 16$ such components in the $N$-$N$ force, which is indeed exemplified
by the 18 components of the recent AV18 potential\cite{AV18}. Dealing with these
complexities, in addition to the 3 continuous coordinates specifying the
positions of 3 nucleons, is a formidable numerical problem.

\begin{figure}[htb]
\epsfig{file=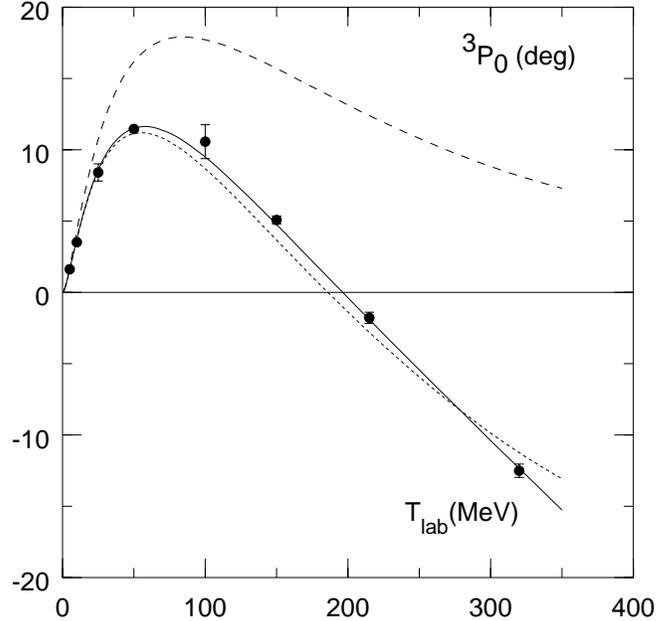,height=3.50in,bbllx=-20pt,bblly=250pt,bburx=648pt,
   bbury=650pt}
  \caption{$^3P_0$ phase shift calculated with OPEP tail for $r > b$   (dashed 
  line), and with either one (dotted) or 3 (solid) short-range interaction 
  terms added.}
\end{figure}

The importance of OPEP is illustrated in Fig.\ (1) from the Nijmegen
group\cite{3P0}. Using a potential that vanishes out to $b=1.4$ fm and
incorporates OPEP plus some two-pion-exchange potential beyond that value leads
to the dashed curve. Clearly, the shape of the phase shift is correct.  Adding a
smooth background contribution from a short-range potential $(r \leq 1.4 \, {\rm
fm})$ produces the dotted curve, while fine tuning leads to the solid curve. 
All of the ``shape'', however, is produced by pion exchange, which is hardly a
surprise given that the pion is the lightest of the mesons exchanged between two
nucleons.

An obvious question is whether a 1-2 MeV uncertainty is a serious handicap in
understanding the physics.  Alternatively, if one wishes to probe the nuclear
force by examining trinucleon properties, what level of calculational accuracy
is a reasonable requirement?  The fundamental problem is determining the
structure of the $N$-$N$ force, and this is impossible to achieve using only the
$N$-$N$ scattering data. Imagine that some $N$-$N$ phase shift is known at all
energies and with infinite accuracy (neither assumption is true), and that there
is no bound state. Under these idealized conditions a potential $V(r)$ (where
$r$ is the separation of the two nucleons) can be deduced that in the
Schr\"{o}dinger equation will reproduce the phase shift. Unfortunately, one can
also deduce a $V(r,p)$ (where $p$ is the relative nucleon momentum) that
reproduces that phase shift equally well.  On-shell (free-nucleon) scattering
cannot produce a unique potential. This led to the idea that making the nucleons
``off-shell'' by placing them in a bound system with a third nucleon might
provide enough additional information to fix the potential, since $V(r)$ and
$V(r,p)$ defined above will definitely produce different tritons.  This is one
aspect of what has become known as the ``off-shell problem''.

We can estimate the uncertainties by noting that the $N$-$N$ system (with
potential $V$) feels the presence of the third nucleon only through the action
of another $V$ and the effect should scale as $V^2$, which has the wrong
dimensions. Another related off-shell problem is that the motion of a pion
propagating between nucleons is conventionally specified only by its transferred
momentum, $\vec{q}$, while its transferred energy, $q_0$, is replaced by other
variables such as $p^2/2M$. This hints that the effective off-shell interaction
scale is set by $\Delta H = V^2/M c^2$, which is correct\cite{offshell} in spite
of the intuitive derivation. Because $V^2$ contains terms linking three nucleons
together and because of the $1/c^2$, this effect is at the same time a
three-body force, an off-shell effect, and a relativistic correction.  Using
reasonable numbers for the triton we estimate $\langle \Delta H \rangle \sim$
0.5-1.0 MeV. Thus the previously noted calculational uncertainties ($\sim$ 1-2
MeV) are unacceptably large, and calculational errors $\lesssim$ 100 keV (which
is approximately 1\% of the binding energy) are required in order to investigate
the three-nucleon effects discussed above. In addition, 1\% absolute experiments
are extremely difficult and uncommon. Consequently, 1\%-error calculations,
known variously as ``exact'', ``complete'', or ``rigorous'', have become the
standard of the field. The ability to achieve this has become our field's major
success story.

\section*{Three-Nucleon Calculations}

The types of problems attacked and the period during which success was
achieved are shown in Fig.\ (2) and Table 1 \cite{BLAST}.
\begin{figure}[htb]
\epsfig{file=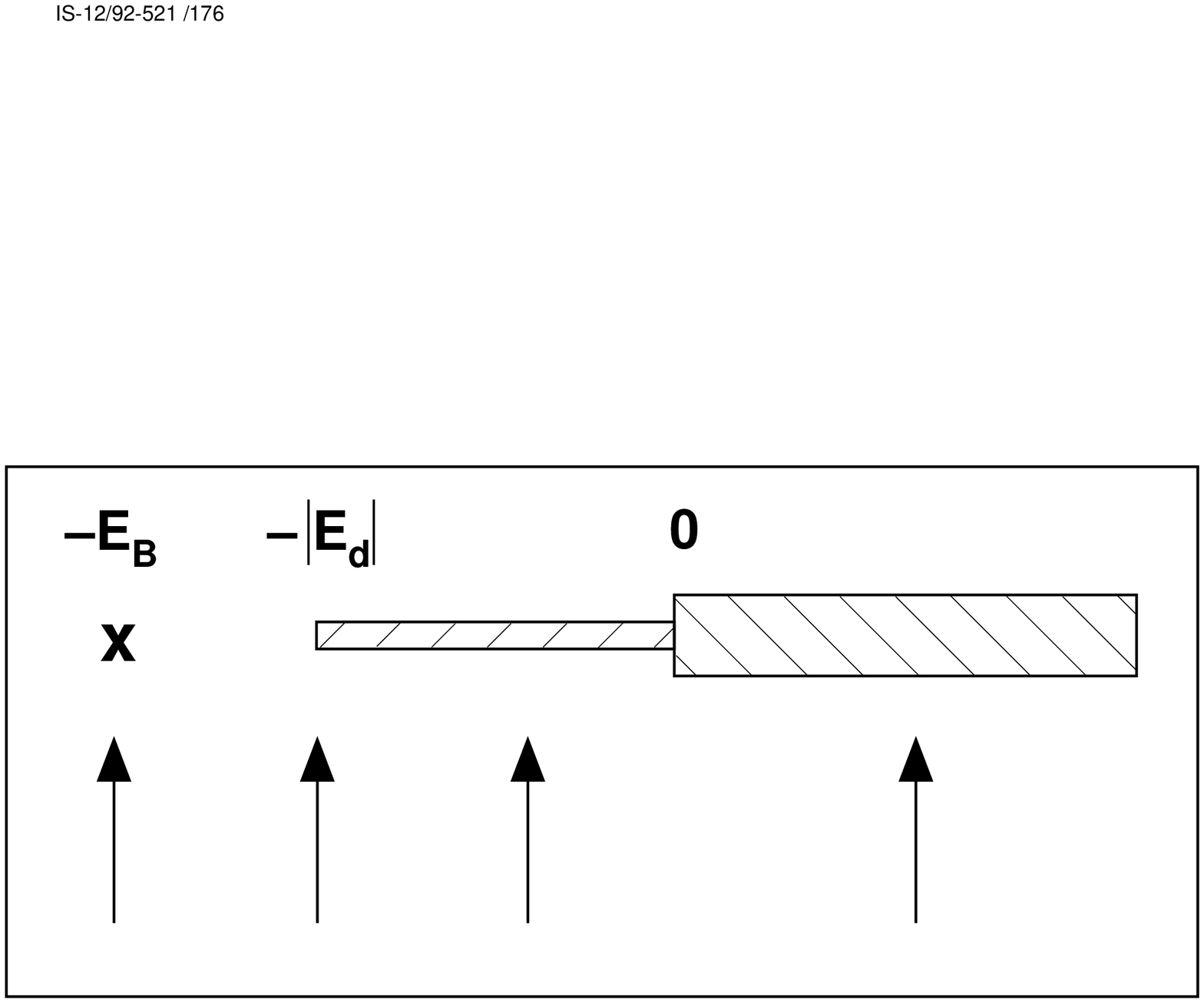,height=1.5in,bbllx=-220pt,bblly=237pt,bburx=570pt,
bbury=495pt,clip=}
\caption{Energy spectrum of $^3$H.}
\end{figure}
There are four regions of energy illustrated in Fig.\ (2) (by arrows) that
conveniently encompass the three-nucleon problem: (1) the trinucleon bound
states (a pole at $-E_B$); (2) zero-energy nucleons scattering from the
deuteron; (3) $N$-$d$ scattering below deuteron-breakup threshold (viz., zero 
total energy); (4) $N$-$d$ scattering above breakup threshold. These problems 
were solved at the 1\% level at times indicated in Table 1. The Los Alamos-Iowa
group was fortunate enough to have participated in half of the entries (top
half) in the table, beginning with the $^3$H bound state in 1985\cite{chen} and
using only $N$-$N$ forces, then adding a 3NF, and finally solving $^3$He in 1987
(which includes a $p$-$p$ Coulomb interaction)\cite{Levin}. Scattering lengths
were calculated a few years later\cite{Levin}. The bound-state problems are
relatively easy, however. Scattering below breakup threshold\cite{Pisa_3} is 
nearly an order of magnitude harder than a bound-state problem, and 
above-breakup scattering is nearly an order of magnitude harder 
still\cite{walter88}. Above-breakup $p$-$d$ scattering is a very recent 
development\cite{Pisa_ab}.

\begin{table}[htb]
\centering
\caption{Complete three-nucleon calculations:
``\bdia''  indicates calculations from mid-late 1980's;
``\bstar'' indicates calculations from the early 1990's;
``$\bullet$'' indicates calculations from early-mid 1990's;
``\bsq'' indicates very recent calculations.}

\begin{tabular}{|l|ccc|}
\hline
Type & NN Force & NN + 3NF & Coulomb \STRUT \\ 
\hline \hline
$E = -E_B$\STRUT   &  \bdia        & \bdia     & \bdia       \\
$E_{Nd} = 0$      &  \bstar       & \bstar    & \bstar      \\
$E < E_{th}$ &  $\bullet$    & $\bullet$ & $\bullet$   \\
$E > E_{th}$ &  \bdia        & $\bullet$ & \bsq        \\ \hline
\end{tabular}
\end{table}

\begin{figure}[htb]
\epsfig{file=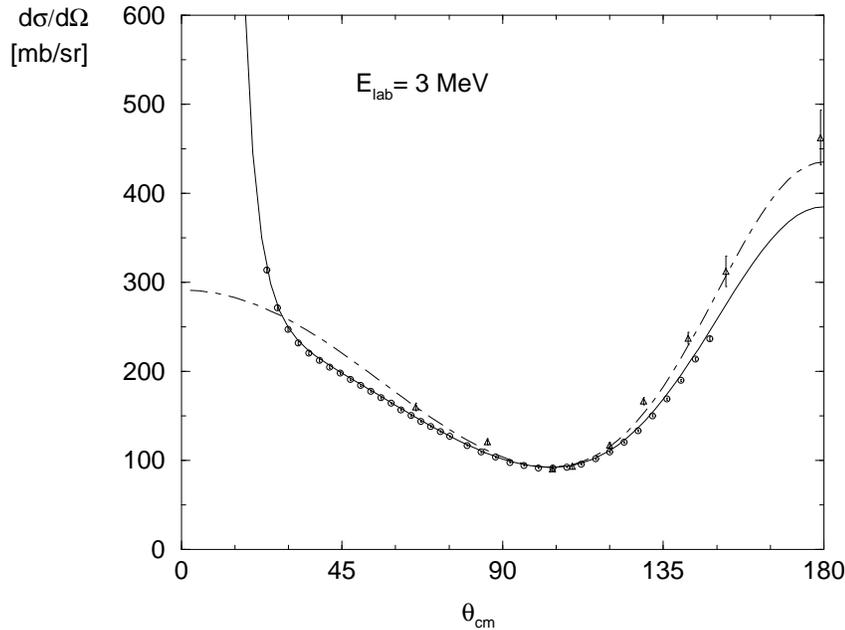,height=3.5in}
\caption{N-d scattering at 3 MeV.}
\end{figure}

A particularly lovely example of this progress is shown in Fig.\ (3), obtained
from the Pisa group\cite{Pisa_3}.  Elastic scattering of 3 MeV nucleons (just
below breakup threshold) from deuterons is calculated and compared to data.  The
solid curve ($p$-$d$) agrees superbly well with the dense, accurate data, while
sparser $n$-$d$ data agree well with the (dashed) calculated values.  Note the
large Coulomb effect at the forward and backward angles.  This plot is rather
typical of differential cross sections:  they are insensitive to the details of
the nuclear force and agree very well with data.  Most spin observables, such as
tensor analyzing powers, also agree well with data.

\begin{figure}[htb]
\epsfig{file=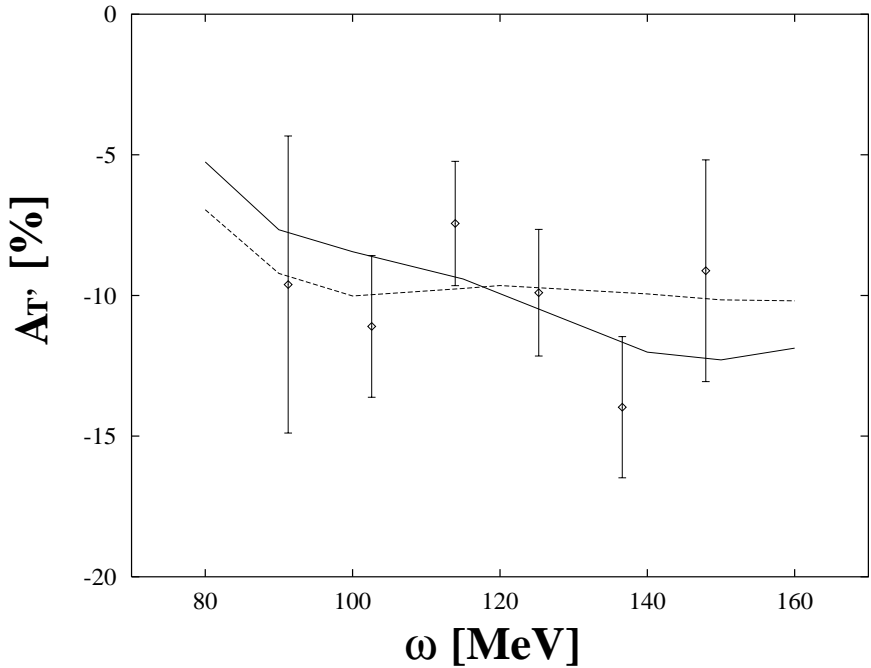,height=3.0in,width=3.5in,angle=-90}
\caption{The spin-dependent asymmetry $A_{T^{\prime}}$ in the reaction
$^3\protect\overrightarrow{\rm He} (\vec{e}, e^\prime n) pp$. The solid curve
depicts the full calculation, while the dashed curve lacks final-state
interactions.}
\end{figure}

Figure (4) shows a very recent calculation \cite{A_T} of an electromagnetic spin
observable, $A_{T^{\prime}}$, in the reaction $^3\overrightarrow{\rm He} 
(\vec{e}, e^\prime n) pp$. The $^3$He target is polarized along the direction of
electron momentum transfer, and the electrons are longitudinally polarized. This
spin-dependent asymmetry in a response function is proportional to $G^n_M$
(neutron magnetic form factor) in the most naive description of the reaction. 
That description is based on the observation that s-waves dominate between the
nucleons in $^3$He.  In that case the two protons are required by the Pauli
principle to have spins anti-aligned, and the entire spin of the nucleus is
carried by the neutron.  The protons do contribute to the reaction because the
tensor force modifies the simple s-wave picture and the protons' spins will be
aligned in D-states, and can contribute to the asymmetry through final-state
$p$-$n$ charge-exchange reactions.  The figure illustrates the Bates
data\cite{gao} compared to two theoretical calculations:  the full calculation
(solid curve) and a calculation (dashed curve) that neglects all final-state
interactions.  The latter calculation would be typical of what was available
until very recently, which illustrates both the difficulty of the calculations
and the progress that has been made.

I would like to summarize this part of my talk as follows:

\begin{itemize}
\item We can now accurately calculate three-nucleon properties.  Most of these
properties, such as differential cross sections and most spin observables (e.g.,
the tensor analyzing power, $T_{22}$\cite{walter}), agree well with data and
depend only weakly on a 3NF. Electromagnetic calculations are very difficult 
and are the state-of-the-art.

\item Spin-isospin degrees of freedom are the biggest impediment to few-nucleon
calculations.

\item Many different techniques are now successfully employed in performing
calculations\cite{joe}.

\item 1\% accuracy is needed in order to disentangle the physics.

\item The most demanding problems drive the progress, and Bates problems are of
this type.

\end{itemize}

\section*{Three-Nucleon Forces}

Three-nucleon forces are small, as we argued earlier for a very special case. In
fact that argument holds for the whole class of such forces, as we shall see. 
If they are so small, are they really necessary, or even interesting? The most
modern potentials produce $^3$H bound states that are underbound by up to 1 MeV.
This defect can be compensated by the addition of a 3NF.  Nevertheless, I do
not consider this to be very compelling evidence for three-nucleon forces.  Are
such forces just ``theorists' toys'' or is there more compelling experimental
evidence?

\begin{figure}[htb]
\epsfig{file=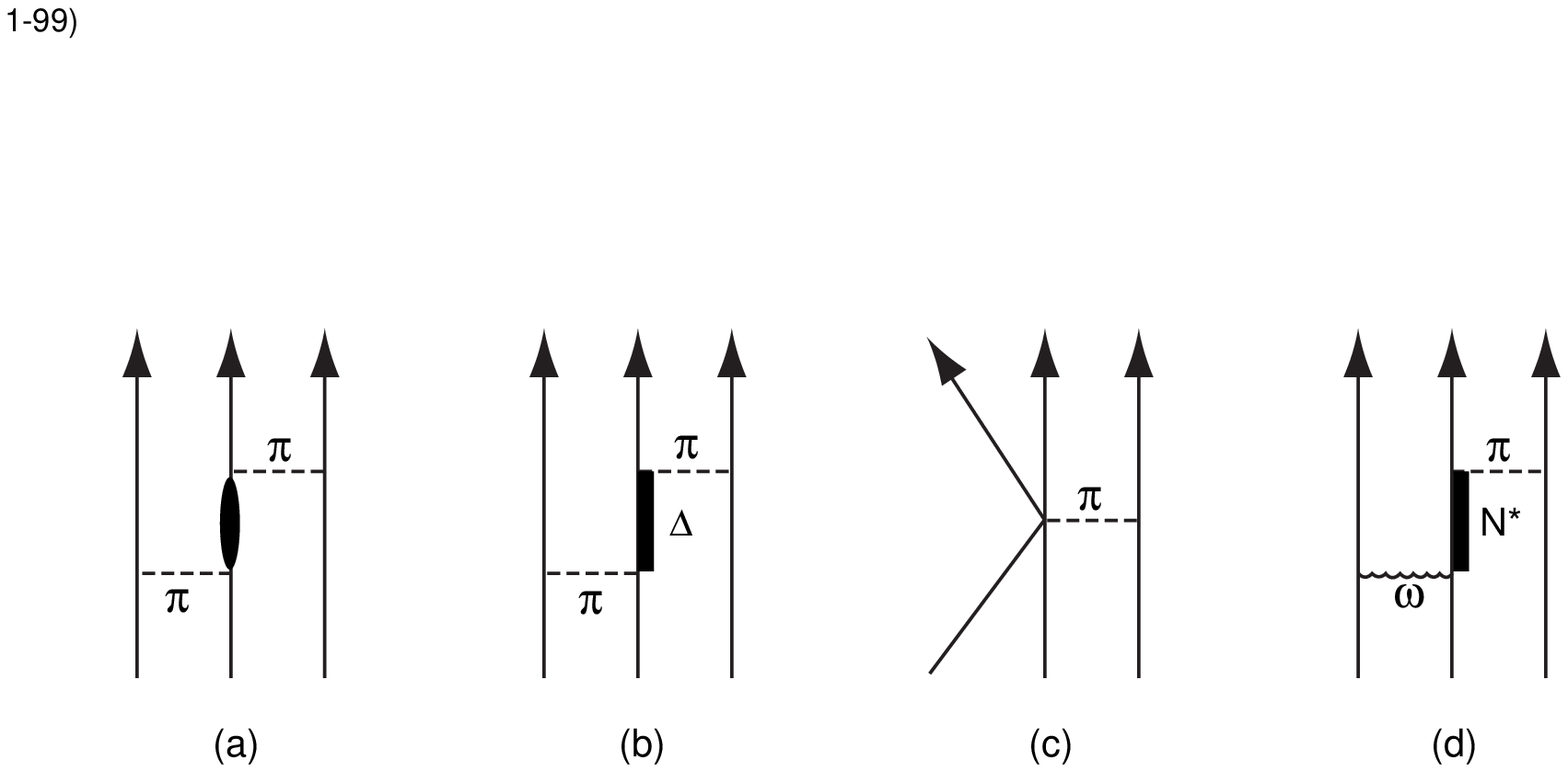,height=1.65in,clip=}
\caption{Mechanisms that contribute to three-nucleon forces. Two-pion-exchange
forces are shown generically in (a), and the important isobar contribution in 
(b). Chiral perturbation theory predicts a large contribution of the type shown 
in (c), a specific mechanism of that type being displayed in (d).}
\end{figure}

In order to answer this question, we must first establish the credentials of the
physics underlying the various models of such forces, which are relatively few
in number. The longest-range mechanisms are those based on 2$\pi$-exchange, and
these have been extensively investigated. Figure (5a) illustrates the generic
force of this type, while Fig.\ (5b) shows the single most important ingredient
(other ingredients are also important). The history of this field is depicted in
Fig.\ (6), a diagram showing the evolution of these forces, all of which are
field-theory based.  Time runs vertically and long lines indicate the oldest
forces.  Near the bottom are the primitive models (PM).  The august
Fujita-Miyazawa model\cite{FM} (FM) is based on $\Delta$-isobars, as is its
offshoot the Urbana-Argonne model\cite{UA} (UA).  To the left are the models
based on chiral symmetry, including the Yang model\cite{Yang} (Y) (the first of
this type, published in 1974) and the Tucson-Melbourne model\cite{TM} (TM), the
oldest such model still in use.  The more recent models based on relativistic
field theories (RFT) are the Brazil\cite{Brazil} (BR) and RuhrPot\cite{RuhrPot}
(RP) models. Finally, the Texas model\cite{texas} (TX) is based on chiral
perturbation theory.  It is clear from this history that the two key ingredients
of 3NFs are:
\begin{itemize}
\item adequate phenomenology (such as isobars).
\item imposing chiral constraints.
\end{itemize}
How does one accomplish this?

\begin{figure}[htb]
\epsfig{file=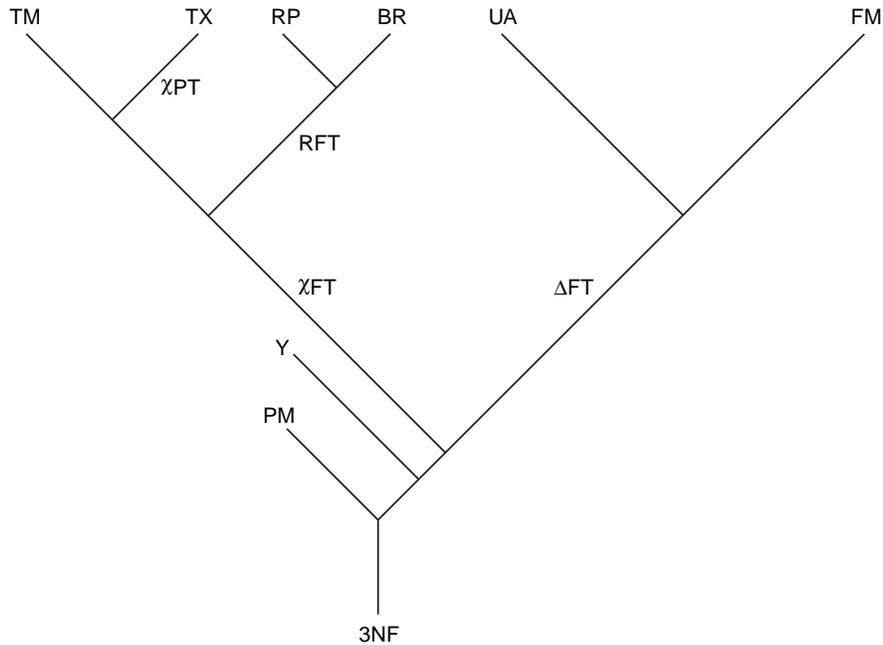,height=4.0in,bbllx=-25pt,bblly=218pt,bburx=547pt,
  bbury=650pt}
\caption{Cladogram \protect\cite{clade} of 2$\pi$-exchange three-nucleon-force
models, showing their history with a vertical time line, together with the
properties that characterized their development.}
\end{figure}

It is believed that the theory underlying the strong interactions is QCD.  The
``natural'' degrees of freedom of this theory are quarks and gluons.  We aren't
required to use these degrees of degrees, however, and traditional nuclear
physics uses effective (observable) degrees of freedom:  nucleons and pions. One
can imagine freezing out all other particles and constructing a theory in this
compressed Hilbert space, in the fashion of (Feshbach) [P,Q] reaction
theory\cite{Feshbach}. Although the resulting operators can be quite
complicated, chiral symmetry, that most important ingredient residing in QCD,
can be implemented in the new theory. This ``QCD in disguise'' is better known
as chiral perturbation theory, and applies to both particles and
nuclei\cite{power}.

Only one aspect of that theory is needed here: dimensional power 
counting\cite{power}. The latter is a kind of (not obvious!) dimensional
analysis based on only two QCD internal energy scales.  The first scale is
$f_\pi$, the pion decay constant ($\sim$ 93 MeV), which controls the Goldstone
bosons and specifically the pion. The second scale is the energy above which we
agree to freeze out all excitations, $\Lambda \sim 1$ GeV, and is the scale
appropriate to the QCD bound states, such as the nucleon, $\rho$ and $\omega$
resonances, etc.  Using these scales, it can be shown\cite{Georgi} that a given
term in a Lagrangian should scale as:
$$
{\cal L}^{(\Delta)} \sim \frac{c}{f_\pi^{\beta} \Lambda^\Delta} ({\rm times \;
various\; fields})\, .
$$
Two important properties are that the power $\Delta$ (used to classify
Lagrangian terms) satisfies $\Delta \geq 0$ (which is a not very obvious
chiral-symmetry constraint), while the dimensionless constant $c$ satisfies
$|c|\sim 1$, the condition of ``naturalness'' (an even less obvious constraint).
Because freezing out degrees of freedom results in effective interactions with
unknown coefficients, the latter condition is the only handle we have on
reasonable values for those constants.

This formal scheme can be implemented in nuclei to estimate the size of various
contributions to potential energies (among others).  An additional nuclear scale
is required, the effective momentum or inverse correlation length, which is
given by $Q \sim m_\pi c$, where $m_\pi$ is the pion mass.  Then it can be shown
that\cite{power}
$$
\langle V_\pi \rangle \sim \frac{Q^3}{f_\pi\Lambda} \sim 30 \
\mathrm{ MeV/pair}\, ,
$$
$$
\langle V_{3NF} \rangle \sim \frac{Q^6}{f_\pi^2\Lambda^3} \sim 1 \ \mathrm{
MeV/triplet}\, .
$$
The latter relationship can also be written as $\langle V_{3NF}\rangle \sim
\langle V_\pi \rangle^2/\Lambda$, which is equivalent to the expression we
developed earlier (since $M \sim \Lambda$) and is also the correct size to
explain the $^3$H binding discrepancy. The use of $\chi$PT is finally leading to
a consensus on 2$\pi$-exchange 3NF terms, and a ``standard'' model of the 3NF is
within reach. All such terms in leading order of $\chi$PT have been calculated,
although some of them have not yet been implemented.

\begin{figure}[htb]
\epsfig{file=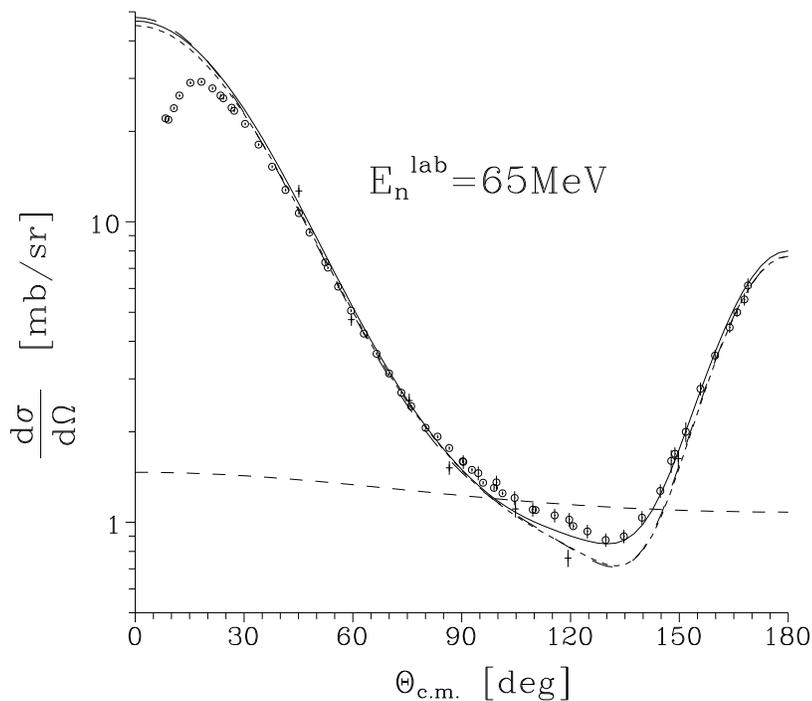,height=4.0in,bbllx=0pt,bblly=0pt,bburx=498pt,
bbury=450pt}
\caption{Differential cross section for 65 MeV proton-deuteron scattering, 
showing calculations with $N$-$N$ forces only (dashed lines), a full calculation
that includes the TM 3NF (solid line), and an estimate of the effect of the 3NF 
alone (long-dashed line).}
\end{figure}

Several of these terms have been checked by testing the tail of the $N$-$N$ 
potential against the set of $p$-$p$ data. That tail is calculated by using the
same Lagrangian building blocks that are used to calculate $\pi$-$N$ scattering
and the 2$\pi$-exchange 3NF. Important elements of the 2$\pi$-exchange $N$-$N$
force were verified\cite{mart}, which validates the corresponding terms in the 
3NF.

In addition to the $^3$H ($^3$He) binding discrepancy, there is one other piece
of experimental evidence for a 3NF that is much stronger. The Sagara discrepancy
\cite{sagara} is illustrated in Fig.\ (7), which shows $p$-$d$ elastic
scattering at 65 MeV. Ignoring the forward direction (where the Coulomb
interaction plays a significant role), the agreement is very good between
calculations with an $N$-$N$ force only (dashed lines) and the experimental data
except in the diffraction minimum. Adding the TM 3NF produces the solid curve,
which is in fairly good agreement with experiment in the minimum.  The small 3NF
effect is depicted by the long-dashed line, which follows from keeping only
those terms linear in the 3NF. This behavior is very reminiscent of Glauber
scattering, with a dominant single-scattering contribution falling rapidly with
angle until the smaller double-scattering term (which has a reduced slope)
becomes significant. This is rather strong evidence for a 3NF, and it persists
to higher energies.

Our final topic is the extension of 3NFs beyond 2$\pi$-exchange.  Chiral
perturbation theory predicts that there are two mechanisms that have pion range
in one pair of nucleons and short range in a second pair, and they should be
comparable in size to the 2$\pi$-exchange mechanisms. The generic force in
$\chi$PT is shown in Fig.\ (5c), and a particular example (the so-called
d$_1$-term) is illustrated in Fig.\ (5d).  All mechanisms affect the $^3$H
binding energy, so this is a poor test of a {\it{specific}} mechanism.  A
tedious examination of low-energy observables\cite{dirk} finds that the
d$_1$-mechanism makes a potentially large contribution to the $n$-$d$ asymmetry,
$A_y$.  This observable at 3 MeV is depicted in Fig.\ (8).  The calculation with
only $N$-$N$ forces is the solid line, which is about 30\% lower than the data. 
The long-dashed curve includes the effect of the TM force, which accounts for 
only about 1/4 of the discrepancy. Adding the d$_1$-term in the 3NF with a
dimensionless coefficient, $c_1 = -1$, produces the short-dashed curve.  The
size and sign of that coefficient are unknown, and the sign was chosen to move
the prediction upward.  Although it appears that a choice of $c_1 = -3$ (and
quite acceptable in size) would resolve the problem, the algorithms used in our
codes failed to converge for such a value, and that final conclusion could not
be checked at the time this manuscript was written.

\begin{figure}[htb]
\epsfig{file=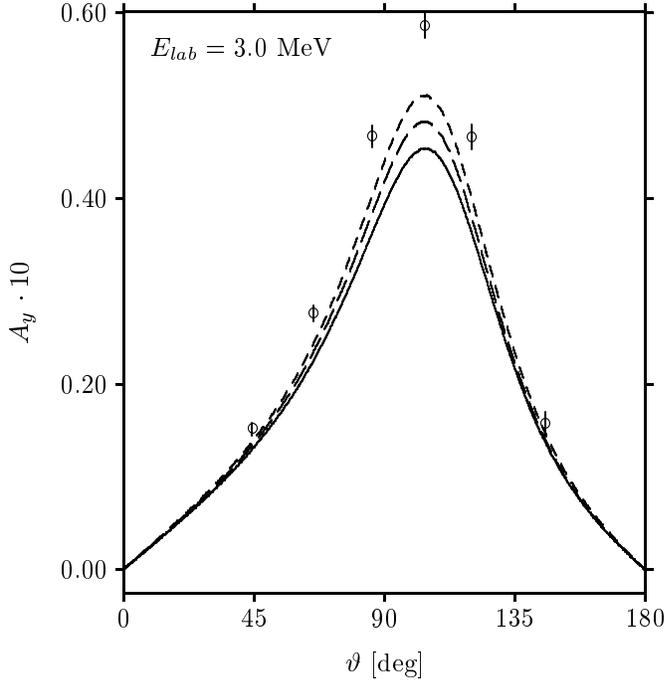,height=4.0in,bbllx=-50pt,bblly=250pt,bburx=550pt,
  bbury=750pt}
\caption{The asymmetry, $A_y$, for 3 MeV neutron-deuteron scattering, calculated
using $N$-$N$ forces only (solid), incorporating the TM force (long-dashed), and
further adding a $d_1$-type force (short-dashed).}
\end{figure}

Nevertheless, it appears that this mechanism could resolve the low-energy $A_y$
puzzle, which has existed for many years and in many forms, for both $p$-$d$ and
$n$-$d$ scattering and in electromagnetic reactions\cite{A_y}. It remains to be
seen whether this mechanism is compatible with the $A > 3$ bound states and 
other data.

We summarize this section as follows.
\begin{itemize}
\item Most three-nucleon observables are insensitive to 3NFs.

\item 3NFs are small in size but appear necessary to reproduce the $^3$H binding
energy, the Sagara discrepancy, and the $A_y$ puzzle.

\item Chiral symmetry provides a unified approach to 3NFs; power counting
identifies dominant mechanisms.

\item The leading-order (dominant) 2$\pi$-exchange 3NFs have been calculated;
they have large isobar contributions.

\item New short-range plus pion-range mechanisms may resolve the low-energy
$A_y$ puzzle.

\item Although much remains to be investigated, a consensus appears to be
developing for the bulk of 3NF terms, and a ``standard model'' of 3NFs may be 
possible in the near future.

\item The basic building blocks of 3NFs have been recently validated by 
verifying the corresponding elements in the tail of the $N$-$N$ potential.

\end{itemize}

\section*{Acknowledgements}

We thank Alejandro Kievsky of the Univ.\ of Pisa and Jacek Golak of the Univ. of
Cracow for providing figures. Walter Gl\"ockle of  Ruhr-Universit\"at Bochum
engaged in a helpful correspondence. The work of J.L.F.\ was performed under the
auspices of the United States Department of Energy.

\end{document}